\documentclass[aps,prd,reprint,superscriptaddress,nofootinbib]{revtex4-2}

\usepackage[colorlinks=true, allcolors=blue]{hyperref}
\usepackage{hypernat}
\usepackage{orcidlink}
\usepackage{bm}
\usepackage{amsfonts}
\usepackage{amsmath}
\usepackage{amssymb}
\usepackage{graphicx}	
\usepackage{booktabs}
\usepackage{diagbox}
\usepackage{makecell}
\usepackage{mathptmx}

\newcommand{\app}[1]{#1}

\newcommand{\rtm}{r_{\mathrm{200m}}}
\newcommand{\rtmd}{\dot{r}_{\mathrm{200m}}}
\newcommand{\Mtm}{M_{\mathrm{200m}}}

\newcommand{\MDM}{M_{\mathrm{DM}}}

\newcommand{\Mtot}{M_{\mathrm{tot}}}
\newcommand{\Mtotd}{\dot{M}_{\mathrm{tot}}}
\newcommand{\Mgas}{M_{\mathrm{gas}}}

\newcommand{\Msol}{\mathrm{M}_{\odot}}
\newcommand{\kpc}{\mathrm{kpc}}
\newcommand{\fgas}{f_{\rm gas}}

\newcommand{\entropunit}{{\rm keV\,cm^{2}}}
\newcommand{\dd}{\mathrm{d}}
\newcommand{\deltaTagn}{\Delta T_{\rm AGN}}

\newcommand{\letter}{PP26L}

\begin{document}

\title{\textbf{The limits of feedback from active galactic nuclei} }%

\author{Andrew Pontzen\orcidlink{0000-0001-9546-3849}}
\email[]{andrew.p.pontzen@durham.ac.uk}
\affiliation{Institute for Computational Cosmology, Department of Physics, Durham University, South Road, Durham, DH1 3LE, UK}

\author{Hiranya V.\ Peiris\orcidlink{0000-0002-2519-584X}}
\email[]{hiranya.peiris@ast.cam.ac.uk}
\affiliation{Institute of Astronomy and Kavli Institute for Cosmology, University of Cambridge, Madingley Road, Cambridge CB3 0HA, UK}
\affiliation{Cavendish Laboratory, Department of Physics, University of Cambridge, JJ Thomson Avenue, Cambridge, CB3 0HE, UK}
\affiliation{The Oskar Klein Centre, Department of Physics, Stockholm University, AlbaNova University Centre, SE 106 91 Stockholm, Sweden}

\author{Joop Schaye\orcidlink{0000-0002-0668-5560}}
\affiliation{Leiden Observatory, Leiden University, PO Box 9513, 2300 RA Leiden, the Netherlands}

\author{Matthieu Schaller\orcidlink{0000-0002-2395-4902}}
\affiliation{Leiden Observatory, Leiden University, PO Box 9513, 2300 RA Leiden, the Netherlands}
\affiliation{Lorentz Institute for Theoretical Physics, Leiden University, PO Box 9506, NL-2300 RA Leiden, The Netherlands}

\date{\today}

\begin{abstract}
We use FLAMINGO to investigate why feedback from active galactic nuclei (AGN) significantly depletes gas in galaxy groups but is ineffective in clusters. We delineate three radial zones: an inner zone where AGN feedback heats halo gas via shocks; an intermediate buoyancy zone where the heated halo gas rises; and an outer zone where the outflow may stall in a termination shock.   Heating in the inner zone self-limits because, once the gas is sufficiently hot, shocks become too weak to deposit further entropy. Consequently, outflows have a `ceiling' entropy value ($360$~and~$370\,\entropunit$ in the FLAMINGO Fiducial m8 and m9 models) that is nearly independent of halo mass. These values (and trends with redshift and feedback variants) are explained using an argument based on the Rankine-Hugoniot relations.    Outflows rise at fixed entropy through the buoyancy zone, escaping the halo if the ceiling value is sufficiently elevated over that of the inflowing gas. This condition is satisfied only for halo masses $\Mtm<10^{13.7}\,\Msol$ (for Fiducial), because inflow entropy tracks the virial relation. Variants with stronger (or weaker) feedback have a higher (or lower) entropy ceiling and a correspondingly modified critical mass of $\Mtm = 10^{14.0}\,\Msol$ (or $10^{13.5}\,\Msol$).  In clusters above the critical mass, the increased inflow entropy causes the outflow to stall and potentially shock at the `splashback' radius.   We derive an expression for the time evolution of the virial gas fraction, which shows how lingering gas is passively reincorporated as the halo virial radius expands. This effect dominates over physical outflows unless they rejoin the Hubble flow; as a result, virial gas fractions rise as a function of mass starting at $\Mtm = 10^{13.0}\,\Msol$, much lower than the critical mass for the transition from outflow to inflow. The combination of these effects explains why groups have depleted gas, while clusters have close to the cosmic baryon fraction.
\end{abstract}

\maketitle

\section{Introduction}

Observations of galaxy groups and clusters with X-ray telescopes have long established that the masses of gas fall short of the cosmic complement of baryons \cite{Vikhlinin06XrayClusters}. However, there is a strong trend with mass \cite{Sun09XrayChandraGroupsClusters,Pratt09GasFractions,Eckert16BaryonFracs}:  gas mass fractions for groups (virial masses $\simeq 10^{13}\,\Msol$) reach as low as $\simeq 0.05$ at small radii, while high-mass clusters ($>10^{14}\,\Msol$) have fractions more comparable to the baryon density in the universe at large ($\simeq 0.15$). Even once the mass locked in stars is taken into account, groups have low total baryon densities, unlike clusters  \cite{Gonzalez13BaryonFractions,Akino22BaryonFractions}. The questions therefore arise: where are the baryons in groups, how did they get hidden from observation, and why do clusters  have a larger baryon fraction?

In recent years, X-ray observations have been joined by powerful complementary probes in the form of Sunyaev-Zel'dovich spectral distortions of the cosmic microwave background  (thermal and kinetic, henceforth tSZ and kSZ \cite{Planck13TSZ,Battaglia17SZHaloConstraints,Bigwood24KSZ,McCarthy25kSZSims}) and fast radio burst (FRB) dispersion measures \cite{Reischke25FRB,Lanman25FRB}. These techniques broadly confirm\footnote{One notable exception is the tSZ power spectrum (as opposed to stacked tSZ). This is found to be in agreement with expectations on large scales, but exhibits a deficit of power inferred on small scales, which remains to be explained \cite{EfstathiouMcCarthy25tSZCl}.} the trends established from X-rays. For example, stacking analyses of tSZ around clusters imply a pressure profile in close agreement with expectations extrapolated from X-rays \cite{Planck13TSZ}. Similarly, kSZ stacks and FRBs confirm strong gas depletion in group-scale halos \cite{Bigwood24KSZ,McCarthy25kSZSims,Hadzhiyska25kSZDESI,LLS25,Reischke25FRB,Lanman25FRB,Qu2026kSZ}. (The strength of the effect is somewhat greater than implied by historical X-ray samples, pointing either to the difficulties inherent in estimating gas fractions \cite{Eckert25kSZXrayConsistency,Seppi26BoundOrBlown,Seppi26ChexMate,McCarthy25kSZSims} or sample selection biases \cite{Eckert11XraySampleBias,Popesso25XrayFgas}.) Moreover, sensitivity to diffuse gas allows kSZ probes to demonstrate that the ``missing'' baryons from groups can be found observationally lying within 2 -- 3 virial radii \cite{Hadzhiyska25MissingBaryonsKsz,Hadzhiyska26Ksz}. This builds on earlier X-ray studies that infer gas fractions increasing with radius \cite{Eckert13GasFractionWithRadius,GhirardiniICMprofiles19}. The questions are therefore sharpened: why do baryons leave the central regions of groups, then accumulate beyond the virial radius? Is this gas reaccreted if a system grows in mass, or is there another reason for the higher gas fractions in clusters?

Energy from a central active galactic nucleus (AGN) has long been the prime suspect for heating and/or removing gas \cite{McNamara07Review,McCarthy10AGNfeedbackInGroups,Fabian12review}. The observed `$M-\sigma$' relation \cite{FerrareseMerritt00Msigma,Gebhardt00MSigma} between the central supermassive black hole mass and stellar velocity dispersions implies the total energy budget is sufficient to heat and unbind a large fraction of baryons \cite{SilkRees98,BoothSchaye10DMBHconnection,Gaspari19SMBHMassRelation}. Multiple simulation suites have reproduced the corresponding relation between black hole mass and stellar mass, alongside mass dependence of gas fractions, provided the highly uncertain physics governing AGN feedback is calibrated appropriately \cite{Schaye10OWLS,Schaye15EAGLE,McCarthy17BAHAMAS,Vogelsberger14Illustris,Pillepich18IllustrisTNG,Dave19SIMBA}. 

One example is the FLAMINGO suite \cite{schaye2023}, which we use in this work because it well reproduces a range of X-ray observational signatures   \cite{Braspenning24FlamingoClusters,Seppi26BoundOrBlown}.  Moreover, FLAMINGO includes variant simulations that systematically alter the effects of AGN feedback. The strength of the feedback in `Fiducial' FLAMINGO simulations is calibrated to reproduce historical X-ray-inferred gas fractions in clusters \cite{Kugel23}, then systematically increased or decreased in the variants. The suite includes further `Jet' variants where the energy is deposited kinetically rather than thermally \cite{Husko22}. 

We previously found that FLAMINGO simulations predict rising gas fractions for virial masses $\gtrsim 10^{13}\,\Msol$  \cite{LLS25,OndaroMallea25}. While this fits qualitatively with the observational picture, it does not provide a physical understanding of why a particular halo mass should emerge at which effects are maximal. For this, we need a theory of why gas leaves groups, and where it stops.

A key observational clue is the observed excess of entropy above self-similar gravitational expectations \cite{Ponman99EntropyExcess}.  If a sufficient entropy excess can be imparted to a significant gas mass, buoyancy then lifts it naturally out of the gravitational potential \cite{Churazov01BuoyantBubblesM87,Voit05Review}. Simulations have shown that much of the high-entropy gas in groups leaves the halo, in agreement with this buoyancy picture \cite{McCarthy11HighEntropyGasLeavesHalos}.

In a companion letter (henceforth \letter), we propose that the overall phenomenology of observations and simulations can be understood if AGN heat gas up to an entropy ceiling that has only a weak dependence on galaxy mass. This naturally leads to a critical halo mass above which virial entropy exceeds the ceiling and feedback becomes ineffective. The argument is particularly complementary to the ``dark nemesis’' work of Bower et al. \cite{Bower17DarkNemesis}, who connect the onset of AGN feedback dominance with the loss of buoyancy of supernova-heated gas. PP26L studies the plot twist: the highest mass galaxy clusters in our universe ultimately overcome the dark nemesis, using the same buoyancy-loss phenomenon.

In this paper, we study the origin and role of buoyancy within the FLAMINGO simulation suite. We quantify gas inflow and outflow rates, their energetics, and the entropy generation as a function of halo mass and radius. We show that the entropy ceiling arises from shock heating, and that the transition from effective to ineffective feedback is well predicted by matching the ceiling to the virial entropy. We find the same overall phenomenology in thermal and kinetic (Jet) feedback implementations, but show in the latter case that the ceiling gains a weak mass dependence due to the change of geometry. 

The structure of the paper is as follows. In Sec.~\ref{sec:methods} we briefly introduce the FLAMINGO simulations and our approach to post-processing. In Sec.~\ref{sec:results} we present our results, starting by outlining the interaction between net gas flows,  dark matter accretion, and virial radius expansion that collectively determine the gas fraction. Sec.~\ref{sec:fgas-non-gas} discusses how the accretion and expansion compete with the net gas flows to select the mass scale at which gas fractions are minimal. Sec.~\ref{sec:gas-flows} examines the overall flows of gas and dark matter as a function of radius. The energetic balance is discussed in Sec.~\ref{sec:outflow-energetics}, and the generation of entropy in Sec.~\ref{sec:entropy-generation}. We introduce the resulting model of outflows in Sec.~\ref{sec:three-zone-model}, based on three radial zones which correspond to inner shock-heating, central buoyant transport, and outer termination or rejoining the Hubble flow. We consider how this phenomenology continues to apply to the kinetic (``Jet") variation of the AGN feedback in Sec.~\ref{sec:jets-variant}. We summarize and discuss the results in Sec.~\ref{sec:conclusions}, where we also comment on how the resulting picture generalizes beyond FLAMINGO.

\section{Methods}\label{sec:methods}

\app{In this Section, we introduce the simulation suite used for our analysis (Sec.~\ref{sec:simulations}), then describe the post-processing undertaken (Sec.~\ref{sec:post-processing}), including halo finding, radial profile generation, and stacking techniques. }

\subsection{Simulations}\label{sec:simulations}

All simulations analyzed are from the FLAMINGO suite, \app{which was recently released to the public \citep{Helly26Flamingo}.  The simulations have} been extensively described \citep{schaye2023}, and here we provide only a brief overview of the most relevant aspects. The simulations are performed in a flat $\Lambda$CDM universe using cosmological parameters based on DES Y3 \cite{Abbott22DESY3Cosmo}; in particular, the Hubble parameter, matter density relative to critical, and baryon density relative to critical take the values  $H_0 = 68.1\,{\rm km\,s^{-1}\,Mpc^{-1}}$, $\Omega_{\rm m,0}=0.306$ and $\Omega_{\rm b,0}=0.0486$ respectively. We primarily analyze the Fiducial m8 simulation, meaning the gas particle mass is $1.34 \times 10^{8}\,\Msol$ (and dark matter particle mass is $7.06 \times 10^8 \, \Msol$). The \app{maximum} gravitational softening length is $2.85\,\mathrm{kpc}$ in physical units. (We adopt physical rather than comoving units throughout this paper.)  We additionally use m9 simulations (gas and dark matter particles respectively $1.07 \times 10^9 \,\Msol$ and $5.65\times10^9 \,\Msol$, with maximum softening $5.70\,\mathrm{kpc}$). We make use of volumes with side length $L=200\,{\rm Mpc}$ which are correspondingly quicker to analyze than the larger $L=1\,{\rm Gpc}$ volumes. None of our results are limited by the available number of halos. We analyzed all available outputs, but focus on results at $z=0.4$, which is a typical redshift where X-ray and SZ constraints on the behavior of intragroup/intracluster gas are available \cite{LLS25}.

At m8 resolution, only simulations with the Fiducial implementation of AGN physics are available, as described below. At m9 resolution we analyze the Fiducial, fgas$+2\sigma$, fgas$-8\sigma$, Jet and Jet fgas$-4\sigma$ variants. The first three use thermal AGN feedback with different calibrations, while the latter two use a kinetic feedback model. 

In both prescriptions, the gas mass accretion rate $\dot{M}_{\rm BH}$ onto supermassive black holes follows a modified Bondi-Hoyle formula (boosted to account for unresolved gas clumping \app{\cite{BoothSchaye09BHFeedback}}). The rate of feedback energy is calculated as $\dot{E}_{\rm AGN} = \epsilon_r \epsilon_f \dot{M}_{\rm BH} \, c^2$ where $\epsilon_r$ accounts for the assumed radiative efficiency of the accretion while $\epsilon_f$ accounts for the fraction of that which couples to gas. The combined efficiency $\epsilon_r \epsilon_f$ is set to $0.015$ in all simulations, determining the available energy budget for AGN feedback to affect halo gas.  

In the Fiducial thermal implementation, feedback energy is accumulated in a subgrid reservoir until it suffices to heat one neighboring gas particle by a temperature increment $\Delta T_\mathrm{AGN}$. The value of $\Delta T_\mathrm{AGN}$ is a parameter, taking values between $10^{7.7}\,\mathrm{K}$ and $10^{8.4}\,\mathrm{K}$ across the variants we analyze. This thresholded heating scheme ensures that each feedback event raises the gas temperature well above that of its surroundings into a regime where cooling is inefficient. In the Jet variants, the same energy budget is instead deposited kinetically: pairs of gas particles near the black hole spin axis are kicked in opposite directions at a calibrated velocity $v_\mathrm{jet}$. After calibration, $v_{\mathrm jet}$ takes either the value $836$ or $1995\,\rm km\,s^{-1}$ for the two Jet variants analyzed here. The spin axis determines the direction of kicks and evolves through gas accretion and black hole mergers. The net effect is that the direction of energy input reorients on timescales long compared to individual Jet episodes, so that the feedback retains a persistent directional character \cite{Husko22}. 

The calibration of these and other numerical parameters is  described in Ref. \cite{Kugel23}. Gaussian process emulators are trained on Latin hypercubes of simulations to probe the parameter dependencies of the $z=0$ stellar mass function and cluster gas fractions $f_\mathrm{gas}$; best fit values to the observed data are then adopted for the Fiducial simulations. The ``fgas$\pm N\sigma$" labels indicate that the observed gas fraction data were shifted by $\pm N$ times the estimated observational uncertainty before calibration, yielding models with systematically stronger or weaker feedback. For example, fgas$-8\sigma$ has $\Delta T_\mathrm{AGN} = 10^{8.4}\,\mathrm{K}$, nearly three times the Fiducial heating temperature, producing substantially lower gas fractions across the group and cluster mass range. This suite of variants allows us to explore how the entropy ceiling and critical mass scale respond to the strength and mode of AGN feedback.

Following conventions of the field, we define the `entropy' of gas to mean its adiabatic constant $p\rho^{-5/3}$. This is stored per-particle in the simulation snapshots, but we renormalize for comparison to observed quantities in fully ionized plasmas as:
\begin{equation}
   K \equiv \frac{p}{\rho^{5/3}} \times \mu \mu_e^{2/3} m_p^{5/3}\,,   \label{eq:K-def}
\end{equation}
where $\mu$ is the mean mass per particle, and $\mu_e$ the mean mass per free electron. We fix these to their values in fully ionized primordial plasma, giving $\mu = 0.59$ and $\mu_e = 1.14$. The resulting quantity is conserved in all adiabatic processes. It also agrees with the observationally conventional definition $K_e=kT/n_e^{2/3}$ in fully ionized regions \citep{Voit05Review}.

\subsection{Post-processing}\label{sec:post-processing}

Halos were identified and tracked across time using HBT-HERONS \citep{ForouharMoreno2025,Han:2017lpe}. This produces a hierarchy of nested, bound halos and subhalos of arbitrary depth. However, we do not make use of the subhalos: during processing, we retain all particles within a spherical region centered on top-level halos only, including those particles that are marked as bound to a subhalo (or as not bound to anything at all). While subhalos host their own galaxies and black holes which contribute to heating halo gas, we find their contribution to the overall energy budget to be small compared with the central. When discussing properties of the AGN, we specifically use the most massive central black hole in the halo, ignoring the subdominant contribution from subhalos' AGN.

Our post-processing pipeline is based on pynbody \cite{pynbody} and tangos \cite{tangos}. Tangos performs calculations in parallel across large numbers of halos, and stores results in a database with an accompanying set of efficient query tools. Black hole accretion rates $\dot{M}_{\rm BH}$ are calculated using the Spherical Overdensity and Aperture Processor \citep[SOAP;][]{McGibbon_2025SOAP} and imported into tangos, to ensure compatibility with other FLAMINGO analyses.  For all other calculations, pynbody provides the underlying I/O and analysis routines. We include only those halos with a dark matter mass exceeding $10^{12.5}\,\Msol$, corresponding to 4\,500 dark matter particles in m8 simulations (and 560 in m9). 

We first calculate halo centres using a `shrinking sphere' approach \citep{Power03}, alongside virial masses $\Mtm$ and radii $\rtm$ satisfying
\begin{equation}
\Mtm \equiv M(\rtm) = \frac{4}{3} \pi \rtm^3 \times 200\, \bar \rho_{\rm m}(z)\,  \label{eq:virial-def}
\end{equation}
where $\bar \rho_{\rm m}(z) = 3 H_0^2 \Omega_{\rm m,0} (1+z)^3 / 8 \pi G$ is the cosmic mean matter density at redshift $z$, and $M(r)$ denotes the total mass enclosed within a sphere of radius~$r$. 

To calculate properties of the flow as a function of radius, we divide particles into 50 radial bins equally spaced in $\log_{10} r$. We calculate all quantities with two binning schemes, corresponding respectively to physical and $\rtm$-scaled radii. The former scheme distributes bins between $50\,\rm kpc$ and $5\,\rm Mpc$; the latter distributes them between $0.05\,\rtm $ and $5.0 \,\rtm$. For our lowest mass halos, $0.05\,\rtm = 17\,\rm kpc$ at redshift $z=0.4$, corresponding to 3 softening lengths in m9 and 6 in m8 simulations. 

In our radial bins, we calculate cumulative masses as a function of radius for dark matter and gas, $\MDM(r_i)$ and $\Mgas(r_i)$, in the obvious way, i.e. by summing the masses of the particles at all radii interior to the bin outer edge. We also calculate binned gas and total density estimates $\rho_{\rm gas}(r_i)$ and $\rho_{\rm tot}(r_i)$ using the particles inside the spherical annulus defined by bin $i$.

Before further processing each halo, we calculate a velocity center using the mass-weighted mean velocity of particles in the central $25\,\rm kpc$ (approximately 4 softening lengths in m9 or 9 softening lengths in m8).  Next, we convert from the  comoving velocities of the simulation output to physical velocities by adding a Hubble recession term $H(z)\, \vec{x}_i$, where $\vec{x}_i$ is the displacement of particle $i$ from the halo center. All particles are then divided into `inflowing' and `outflowing' classes, based on whether their radial velocity $v_{r,i}$ with respect to the halo center is $< 0$ or $\ge 0$ respectively. Total inflow $\dot{M}_{\rm in}(r)$ and outflow $\dot{M}_{\rm out}(r)$ profiles are estimated by summing $m_i v_{r,i} / \Delta r$ for each inflowing/outflowing particle in each radial bin (where $m_i$ is the particle mass and $\Delta r$ the radial bin width). Inflow and outflow energy profiles are constructed similarly, based on the thermal enthalpy and kinetic energy of each particle.

The thermodynamic properties of outflowing and inflowing gas are calculated weighted by the flow rate, which appropriately down-weights the contribution from any near-static parcels of gas.  For example, when calculating the entropy profile of outflowing gas for a halo, we weight each particle's contribution by $\rho v_r$ for that particle, then weight each halo's contribution to the final stack by $\dot{M}_{\rm out}(r)$. This weighting scheme ensures that we capture a representative measure of the thermodynamic state of gas actually contributing to the outflows (or inflows), even if they are highly episodic and/or anisotropic. 

Errors on stacked flow and thermodynamic profiles are estimated using a bootstrap technique. Specifically, we resample 1000 times with replacement from the population of halos in each mass bin, and recalculate mean quantities independently in each resampled stack. The $(16, 84)\%$ range  of resulting estimates is adopted as the uncertainty on our stacked quantities, and shown throughout all figures as shaded bands.

\begin{figure}
    \centering
    \includegraphics[width=1.0\linewidth]{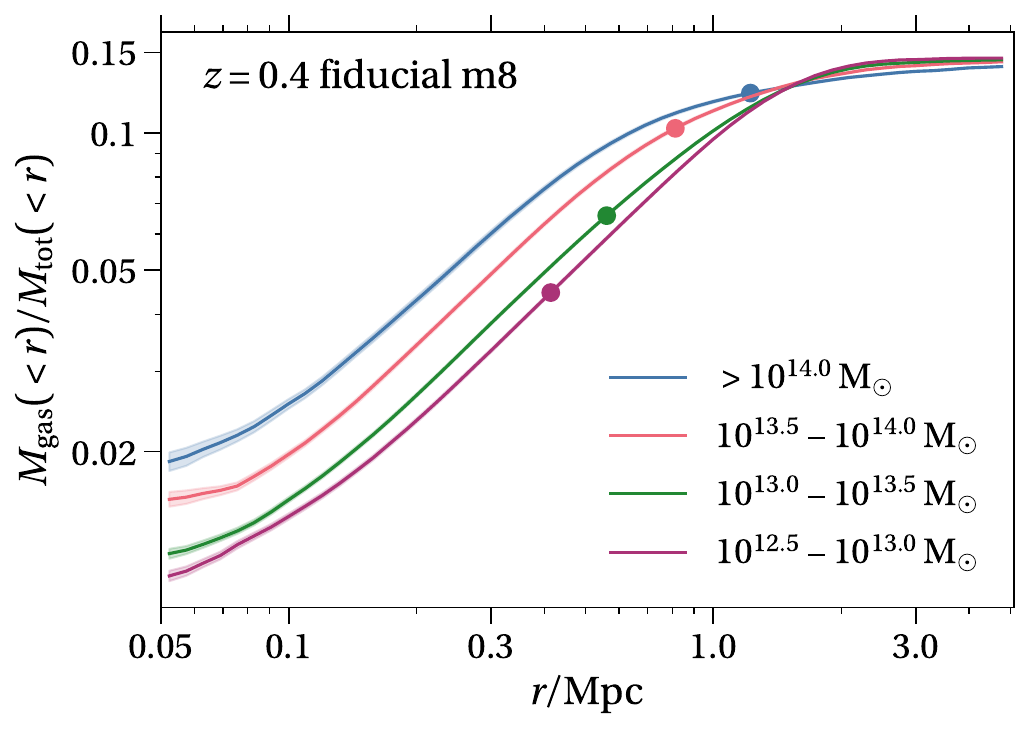}
    \caption{The gas-to-\app{total} ratio as a function of physical radius measured in the Fiducial m8 simulation at $z=0.4$, binned by $M_{200m}$ in 0.5 dex bins starting at $10^{12.5}\Msol$ (purple, green, pink and blue lines respectively). The mean value of $r_{200m}$ in each bin, and the corresponding value of the gas fraction, is indicated as a point. Evaluating at $r_{200m}$ shows a greater mass-dependence of gas fractions than evaluating at a fixed physical radius.  }
    \label{fig:r200m-scaling}
\end{figure}

\section{Results}
\label{sec:results}

Fig.~\ref{fig:r200m-scaling} shows the stacked gas fractions enclosed within a physical radius $r$, $\Mgas(r)/\Mtot(r)$ in the m8 Fiducial simulation at $z=0.4$. The four lines from bottom to top  show bins of $\Mtm$, respectively $10^{12.5} - 10^{13.0}\, \Msol$ (purple), $10^{13.0} - 10^{13.5}\, \Msol$ (green), $10^{13.5} - 10^{14.0}\, \Msol$ (pink) and $>10^{14.0}\, \Msol$ (blue). Shaded bands indicate the bootstrapped errors on the mean, which are small compared with the effects that we will consider. The point on each line indicates the mean location of $\rtm$ for that mass bin. From this, we conclude that the gas fraction within  $\rtm$  is a steeply increasing function of $\Mtm$ above $10^{13.0}\,\Msol$, in agreement with Ref.~\cite{LLS25} (see in particular fig. 5 there). 

Fig.~\ref{fig:r200m-scaling} further illustrates that the gas depletion within the halo virial radius is controlled by \app{three factors}:

\begin{enumerate}
\item The physical gas inflows and outflows which determine the numerator of the ratio plotted;
\item The flows of other material, primarily dark matter, which determine the denominator of the ratio;
\item The differing $\rtm$, which affects where the curve is evaluated, and which in turn arises from the differing total mass of the halo.
\end{enumerate}

While the physics of gas outflows is the main subject of this paper, nonetheless effects 2 and 3 have equally important impacts on measured gas fractions, and compound the difficulties associated with attaining low gas fractions in high-mass clusters. We therefore begin by quantifying these contributions, before moving on to focus on the gas dynamics which determine contribution 1.

\begin{figure}
    \centering
    \includegraphics[width=1.0\linewidth]{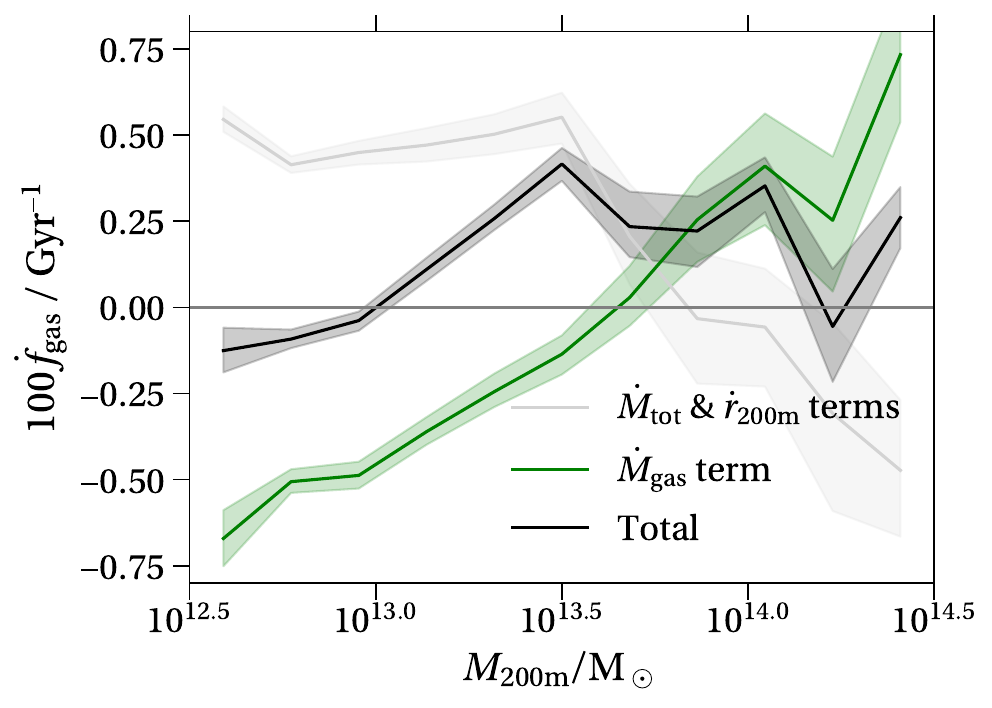}
    \caption{The rate of change in the gas-to-total ratio at $\rtm$ measured in the Fiducial m8 simulation at $z=0.4$, binned by $\Mtm$. \app{The black lines show the total evolution, calculated using Eq.~\eqref{eq:fgas-dot}. Green and gray lines respectively show the contribution from the first term (the physical gas flows) and from the remaining two (combining the effects of the total flow rate, and the changing virial radius as a function of time).} While gas outflows occur from all halos with $\Mtm <10^{13.7} \Msol$, virial gas fractions only decline in halos with  $\Mtm<10^{13.0}\,\Msol$. This is due to the expansion of $\rtm$ into increasingly gas-rich material.}
    \label{fig:fgas-evolution-breakdown}
\end{figure}

\subsection{Non-gaseous contributions to $f_{\rm gas}$}\label{sec:fgas-non-gas}

We start by defining the virial gas fraction as
\begin{equation}
f_{\rm gas} \equiv \frac{\Mgas(\rtm)}{\Mtot(\rtm)}\,.
\end{equation}
Evaluating the time derivative of this ratio shows how the gas fraction of a given halo changes with time as it grows and evolves. We find that
\begin{equation}
    \dot{f}_{\rm gas} =  \left. \frac{\dot{M}_{\rm gas}}{\Mtot} \right|_{\rtm} - f_{\rm gas}  \left. \frac{\dot{M}_{\rm tot}}{\Mtot} \right|_{\rtm} + \frac{\dd f_{\rm gas}}{\dd r} \dot{r}_{200\rm m}\,.\label{eq:fgas-dot}
\end{equation}
The three terms in order correspond precisely to (1) -- (3) in the list of effects given above. We will refer to the first as the gaseous term, and the latter two as non-gaseous. While of course gas is involved in all three terms, it is primarily the evolution of dark matter which controls the latter two.

To evaluate Eq.~\eqref{eq:fgas-dot} numerically requires estimates of $\dd \fgas / \dd r$ and $\rtmd$. The former is obtained from binned density estimates of the gas ($\rho_{\rm gas}$) and total density ($\rho_{\rm tot}$) at $\rtm$, yielding 
\begin{equation}
\frac{\dd \fgas}{\dd r} = \frac{4 \pi \rtm^2}{\Mtm} \left( \rho_{\rm gas} - \fgas \rho_{\rm tot} \right)\,.
\end{equation}
An expression for $\rtmd$ can be obtained by taking the time derivative of the definition~\eqref{eq:virial-def}, giving
\begin{equation}
\rtmd = \frac{\Mtotd(\rtm) - \Mtm \,\dot{\bar \rho}_{\rm m}/\bar \rho_{\rm m} }{ 3 \Mtm / \rtm - 4 \pi \rtm^2 \rho_{\rm tot}(\rtm)}\,.
\end{equation}
This differential expression includes both physical growth (the $\Mtotd(\rtm)$ term) and pseudo-evolution (the $\Mtm \,\dot{\bar \rho}_{\rm m}/\bar \rho_{\rm m}$ term), in a similar way to the integral expressions previously derived by Ref. \cite{Diemer13}. The $\Mtotd$ term is evaluated using our approach to mass flux described in Sec.~\ref{sec:post-processing}, while $\dot{\bar{\rho}}_{\rm m}/\bar{\rho}_{\rm m} =-3H(z)$. 

Using the above expressions, we calculate $\dot{f}_{\rm gas}$ for every halo.  Figure~\ref{fig:fgas-evolution-breakdown} shows the mean $\dot{f}_{\rm gas}$, binned by $\Mtm$, as a black line. The grey band shows the bootstrapped uncertainties. While this result is calculated at $z=0.4$, we verified that its qualitative behavior is the same across cosmic time.  Virial gas fractions increase ($\dot{f}_{\rm gas}>0$) for all halos with $\Mtm > 10^{13}\,\Msol$, explaining why FLAMINGO generates an upturn in $f_{\rm gas}$ starting at this threshold mass \cite{LLS25}. 

However net physical gas outflows persist to much higher masses $\Mtm \simeq 10^{13.7}\,\Msol$. This is illustrated by the green line which includes only the gaseous term in Eq.~\eqref{eq:fgas-dot}. In the interval between these two thresholds, i.e. $10^{13}\,\Msol < \Mtm < 10^{13.7}\,\Msol$, the positive $\dot{f}_{\rm gas}$ arises from the gas-rich material outside the virial radius; as this becomes incorporated into the somewhat arbitrary boundary given by $\rtm$, the gas fraction rises. The physical outflow is present, but not strong enough to prevent this essentially passive growth in $f_{\rm gas}$.

The gray points include the non-gaseous terms (i.e. the last two) in Eq.~\eqref{eq:fgas-dot}. The rapid growth of $\rtm$ (the positive third term) overcomes the growing $M_{\rm tot}$ (the negative second term). Consequently, the combined effect of these non-gaseous terms push $f_{\rm gas}$ back towards the cosmic mean. Even where gas outflows are successfully lowering $f_{\rm gas}$ (i.e. for $\Mtm<10^{13.0}\,\Msol$), the rate at which $f_{\rm gas}$ declines is significantly slowed by the non-gaseous terms due to the near-cancellation between the gaseous and non-gaseous terms in Eq.~\eqref{eq:fgas-dot}.

At very high mass ($\Mtm > 10^{14.3}\,\Msol$) an opposite cancellation between gaseous and non-gaseous terms takes place.  The green line indicates a strong net inflow of gas. However, this is balanced by a similar inflow of dark matter, giving overall no net evolution in $\fgas$. Another way to understand this effect is that $\fgas$ is already close to the cosmic mean value at such high masses, and there is no physical mechanism via which $\fgas$ can be raised above this ceiling.

In summary, three effects combine to determine $f_{\rm gas}$, and in this section we showed that all three play important roles. However, even the two terms that do not feature $\dot M_{\rm gas}$ are not truly independent of the physical gas flows. For example, the net effect of the gradually expanding $\rtm$ depends on how far any gas that left the halo has propagated into the universe beyond. We next investigate more closely why the physical gas flows turn from net outflow to inflow above $10^{13.7}\,\Msol$, and what happens to the gas that leaves halos.

\subsection{Gas flow rates}\label{sec:gas-flows}

\begin{figure}
    \centering
    \includegraphics[width=1.0\linewidth]{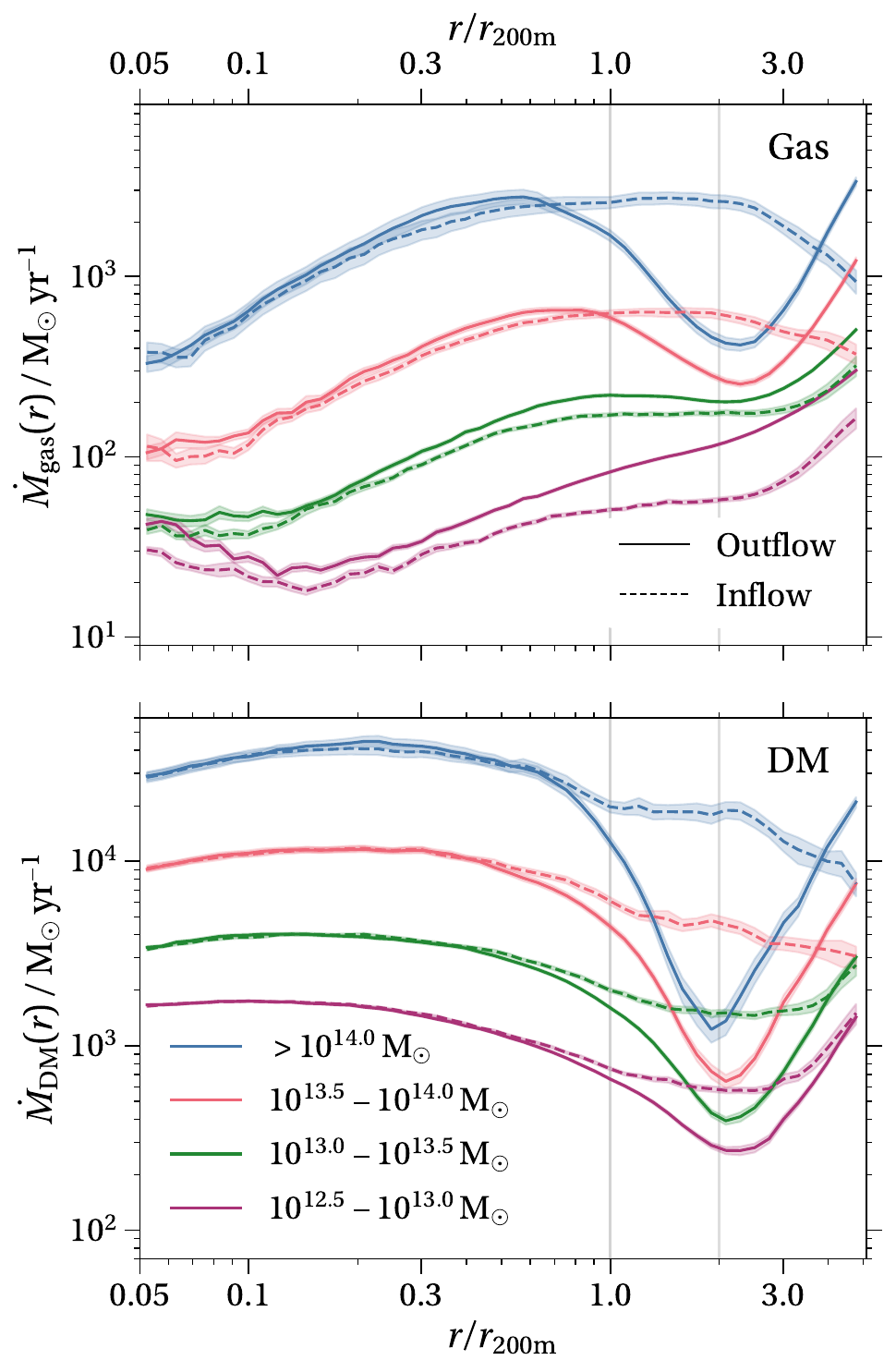}
    \caption{The average outflow (solid) and inflow (dashed) rates as a function of radius normalized to $\rtm$. The top panel shows gas flows, while the bottom panel shows dark matter flows. Colors indicate $\Mtm$ bins, as in Fig.~\ref{fig:r200m-scaling}. Gas and dark matter alike show approximate equilibrium between in- and outflow deep within halos. Dark matter inflows always exceed outflows at the virial radius (leading to net halo growth), but in halos below the threshold $\simeq 10^{13.7}\,\Msol$, gas outflows exceed inflows at the virial radius and beyond.  The upturn in outflows beyond the splashback radius ($\simeq 2\rtm$) is generated by the Hubble flow. \app{Vertical lines show $\rtm$ and $2\rtm$ for reference.}}
    \label{fig:outflow-inflow-profiles}
\end{figure}

\begin{figure}
    \centering
    \includegraphics[width=1.0\linewidth]{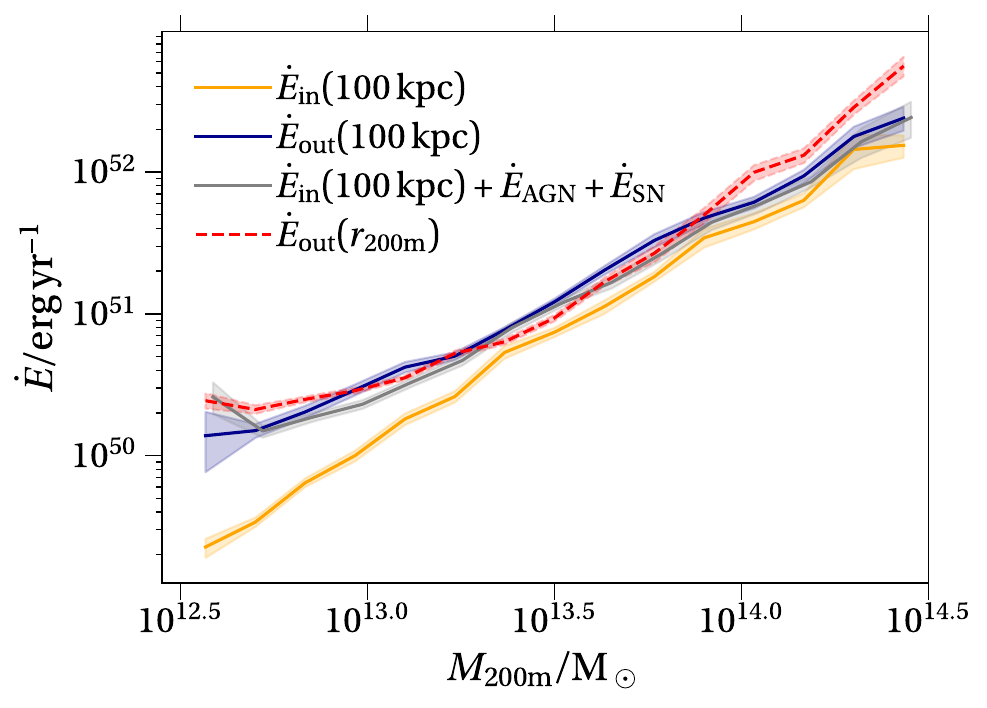}
    \caption{The non-gravitational energy flux (enthalpy + kinetic energy) in gas inflows (orange line) and outflows (blue) measured at a fixed physical radius of $r=100\,\kpc$, as a function of $\Mtm$. The energy in the inflows added to the \app{supernovae and} AGN feedback energy is shown by grey line. This agrees well with the outflow energies (blue). The energy is then conserved out to the virial radius (red dashed line), indicating that the gas is buoyant and so does not lose energy to the gravitational potential.}
    \label{fig:energy-inflow-outflow-vs-mass}
\end{figure}

Fig.~\ref{fig:outflow-inflow-profiles} shows the rate at which gas flows out of (solid) and into (dashed) halos as a function of radius, normalized to $\rtm$, in the same $\Mtm$ bins as used for Fig.~\ref{fig:r200m-scaling}. The top panel shows the gas flows of primary interest here, while the lower panel shows dark matter flows for comparison. Focusing first on small radii ($\lesssim 0.5\, \rtm$), there is an approximate equilibrium between inflow and outflow for both DM and gas. In the case of DM, the mass flow is close to constant as a function of radius. The innermost radii act as an effective reflecting barrier, where inflowing particles reach pericenter and return as an `outflow' on the outgoing part of their orbit. In equilibrium this enforces the condition $\dot{M}_{\rm DM, out}(r) \simeq \dot{M}_{\rm DM, in}(r)$. This near-equality remains true in the gas component too; however there is a slight excess in the gas outflow rate over the inflow rate, which we will shortly show is generated by AGN activity.  

At larger radii DM outflows progressively stall, reaching a minimum at $\simeq 2 \rtm$. This reflects the distribution of apocenters for bound DM particles which drops rapidly between $\rtm$ and $2 \rtm$, giving rise to the well-studied `splashback' feature in the density profile \citep{Adhikari14, MoreDiemerKravtsov15}; see also Ref.~\cite{Towler24GasSplashback} for a discussion of the properties of gas at the splashback radius of FLAMINGO halos. Beyond the splashback feature, the outflows rise again; this is due to the Hubble flow carrying material away.

This same splashback feature can be seen in the gas flows for halos exceeding the critical mass, $\Mtm>10^{13.7}\,\Msol$. Outflowing gas in this mass regime remains predominantly bound.  However, in lower mass halos the gas splashback feature is absent, with the outflows instead smoothly matching onto the Hubble flow, enabling gas to escape to large radii. In the intermediate mass regime, $10^{13.0}<\Mtm/\Msol<10^{13.7}$, gas does escape from the virial radius but we conclude that it lingers just outside. This is because at $r \simeq \rtm$ in this mass bin (green lines), the outflow and inflow rates are constant with radius. Despite a slight net outflow, the density of gas $\rho_{\rm gas}$ then remains constant with time, since
\begin{equation}
\dot{\rho}_{\rm gas}(r) = \frac{1}{4 \pi r^2}\left( \frac{\dd \dot{M}_{\rm in}}{\dd r} - \frac{\dd \dot{M}_{\rm out}}{\dd r} \right) \simeq 0\,.
\end{equation}
This `lingering' effect, where gas densities at the virial radius do not decline over time despite the presence of net outflows, explains why the non-gaseous terms have such a pronounced effect. Even while the gas density inside $\rtm$ is dropping, the gas density just outside $\rtm$ remains fixed. As $\rtm$ then expands, the gas fraction quickly rises towards the cosmic mean, as seen  in Figs.~\ref{fig:r200m-scaling} and~\ref{fig:fgas-evolution-breakdown}.

\subsection{Energetics of outflows}\label{sec:outflow-energetics}

Outflows require a source of energy to power them, and the obvious suspect is feedback, primarily from AGN. To investigate, we start by calculating rates of energy inflow and outflow. \app{Gas crossing a shell advects kinetic and thermal energy, and its pressure $p$ additionally performs work at a rate per unit area of $p v_r$. The latter two terms are combined into a single enthalpy}, giving a total flow rate (temporarily neglecting gravity) of
\begin{equation}
    \dot{E}_{\rm in / out} \equiv \frac{1}{\Delta r} \sum_i m_i \left|v_{r,i}\right| \left(\frac{|\vec{v}|_i^2}{2} + \frac{5}{2} \frac{k T_i}{\mu m_p} \right)\,,
\end{equation}
where the sum over $i$ includes all inflowing or outflowing particles in the shell of width $\Delta r$.  We stack the energy profiles in halo mass bins without further weighting, so the results can be used to diagnose the ensemble-averaged energy balance interior to the shell. \app{If the population is in a state of quasi-equilibrium, where on average the energy interior to a given shell at radius $r$ is evolving slowly \cite{SharmaTheuns20Ikea}, we expect 
\begin{equation}
  \dot{E}_{\rm out} - \dot{E}_{\rm in} \simeq  \dot{E}_{\rm AGN} + \dot{E}_{\rm SN} - \dot{E}_{\rm cool}\, + \Phi(r) (\dot{M}_{\rm in} - \dot{M}_{\rm out})\,,
\end{equation}
from integrating the hydrodynamical energy conservation equation over the volume interior to $r$. Here $\dot{E}_{\rm AGN}$ and $\dot{E}_{\rm SN}$ are the interior rate of AGN and supernova feedback energy injection per unit time, and $\dot{E}_{\rm cool}$ the total interior cooling rate. Any given halo may not satisfy this detailed balance, but averaged over the population it must hold unless there is dramatic evolution with redshift.}

Fig.~\ref{fig:energy-inflow-outflow-vs-mass} shows the resulting picture of energy balance at $r=100\,\kpc$ (orange and blue lines for inflow and outflow respectively). The grey line shows the inflow energy rate added to the feedback energy deposition rate from AGN and supernovae in the central galaxy. \app{We calculate based on the rate at which energy is injected in FLAMINGO, which has a built-in correction for subgrid radiative losses.} The supernova feedback is strongly subdominant, although not entirely negligible at group scale\footnote{For example, we calculate the supernova energy input to groups with $\Mtm = 10^{13}\,\Msol$ to be $2 \times 10^{49}\,{\rm erg\,yr^{-1}}$, strongly subdominant to that contributed by AGN ($10^{50}\,{\rm erg\,yr^{-1}}$).}. The inflow + feedback energy budget (grey line) agrees well with the outflow energy rate (blue line), demonstrating that a simple model in which inflowing material has its energy `boosted' by AGN feedback before returning in an outflow phase works well for the FLAMINGO physics model. 

By additionally measuring the outflow energy flux through $\rtm$ (red dashed line) we find approximate conservation of thermal + kinetic energy throughout the outflow.  This three-way agreement (between AGN-boosted central inflow energy, central outflow energy, and virial outflow energy) can be found in all simulation variants that we analyzed, despite the fact that we have neglected radiative losses, feedback energy injection from satellites, and the effect of the gravitational potential. We comment on these in turn.

Addressing first \app{the resolved} radiative losses in the central 100 kpc,  Ref.~\cite{Braspenning24FlamingoClusters} previously found a wide range of cooling times in the inner 100 kpc of FLAMINGO clusters, in agreement with the diversity of cooling rates seen in observations. As we will see shortly, the cooling has the effect of decreasing the entropy of inflowing gas. However, its overall effects on the energetic balance must be small, given the close agreement between $\dot{E}_{\rm out}$ and $\dot{E}_{\rm in}+\dot{E}_{\rm AGN}+\dot{E}_{\rm SN}$, and the lack of any other sources of energy. Cooling rates at the centres of real groups and clusters are presumably inflated by additional small-scale, dense gas that is not resolved in FLAMINGO, but \app{as noted above, this is} absorbed into $\epsilon_f$. We similarly find the energetic contribution from satellites to be small, because the gas supply reaching their central galaxies and AGN to power feedback is much smaller than that of the central. 

We finally consider the potential energy of gas. Only potential differences carry a physical meaning; consequently, potential energy plays no role in a  direct comparison between $\dot{E}_{\rm out}$ and $\dot{E}_{\rm in}$ at a fixed 100 kpc radius. However when comparing $\dot{E}_{\rm out}$ between 100 kpc and $\rtm$ (blue and red dashed lines respectively in Fig.~\ref{fig:energy-inflow-outflow-vs-mass}), considering the gravitational potential becomes vital. 

For example, at $\Mtm=10^{13}\,\Msol$, \app{the simulation's stacked} potential difference between $100\,\kpc$ and $\rtm$ is $\Delta \Phi = 4.8 \times 10^{48}\,{\rm erg}\,\Msol^{-1}$. Removing gas from $100\,\kpc$ at the measured rate ($\dot{M}_{\rm out} = 41 \,\Msol\,{\rm yr}^{-1}$) and having it reach $\rtm$ would therefore require $2.0\times 10^{50}\,{\rm erg}\,{\rm yr}^{-1}$, comparable to the thermal and kinetic energy in the outflow ($3.2 \times 10^{50}\,{\rm erg\,yr^{-1}}$). That none of this energy is lost to the potential during the journey implies the AGN itself is only part of the picture.

This can be explained if the outflowing gas is buoyant, in which case the gravitational dynamics pumps energy from inflows to outflows. The inflow mass rate of gas (and therefore the rate of potential energy flux) is somewhat lower than but still comparable to the outflow rate ($36\,\Msol\,{\rm yr}^{-1}$ for the inflows reaching 100\,{\rm kpc}). This provides the balancing reservoir of energy to make sense of the outflow.  The key physics controlling an outflow's ability to escape a halo is therefore its buoyancy and, by extension, entropy rather than energy. 

\subsection{Entropy generation}\label{sec:entropy-generation}

\begin{figure}
    \centering
    \includegraphics[width=1.0\linewidth]{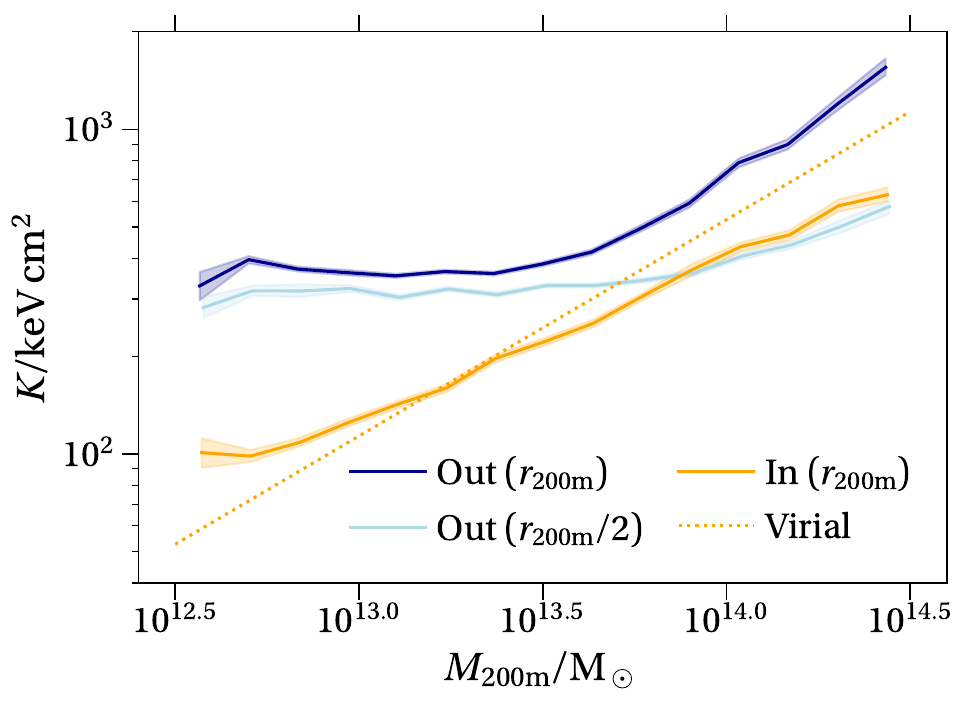}
    \caption{The entropy in gas inflows and outflows at the virial radius (orange and blue lines) as a function of $\Mtm$. The entropy of inflows roughly follows the virial relation (dotted orange line), while the entropy of outflows plateaus at $360\,\entropunit$ in the outflowing regime ($\Mtm \lesssim 10^{13.7}\,\Msol$). Outflow entropies at smaller radii are plotted as a light blue line ($r=\rtm/2$), revealing that the mass-independent plateau extends to much higher masses inside the halo. }
    \label{fig:entropy-inflow-outflow}
\end{figure}

The preceding study of energy concluded that FLAMINGO's AGN-driven gas outflows must be buoyant. This is confirmed in Figure~\ref{fig:entropy-inflow-outflow}, which shows the measured entropy of inflows and outflows at $\rtm$ (orange and dark blue respectively).  We also plot as a dotted line the virial entropy, defined by
\begin{equation}
    K_{\rm 200m}(\Mtm, z) = \mu \mu_e^{2/3} m_p^{5/3} \left( \frac{\pi^2 G^4 \Mtm^2 \Omega_{\rm m,0}}{450 H_0^2 (1+z)^3 \Omega_{\rm b,0}^2}\right)^{1/3}\,. 
\end{equation}
Inflows at $\rtm$ follow approximately the virial relation. However, as expected by the energy arguments, outflows at $\rtm$ show elevated entropy. In particular, they exhibit a flat plateau at $360\,\entropunit$ for halos in the mass regime $\Mtm < 10^{13.7}\,\Msol$.  For reasons that will become clear momentarily, we refer to this plateau as a `ceiling' for AGN entropy production. 

In the Fiducial thermal feedback employed, gas is heated\footnote{In principle $\deltaTagn$ is a {\it minimum} temperature for a particle heated by an AGN feedback event. In practice, we find that 95\% of feedback-affected particles attained maximal temperatures $T_{\rm max}$ in the narrow interval $\deltaTagn < T_{\rm max} < 1.2 \deltaTagn$ \app{for the mass range covered here}. } to $T \simeq \deltaTagn \simeq 10^8{\rm K}$ and the central resolved density is $\simeq 0.1\,{\rm cm^{-3}}$, both independent of halo mass. This does not account for the ceiling: inserting the density and temperature values into Eq.~\eqref{eq:K-def} yields $K_{\rm AGN} \simeq 40\,\entropunit$, an order of magnitude too small. \app{We additionally verified that the total mass of gas heated directly by the AGN subgrid is an order of magnitude  smaller than the outflow rates. We conclude that buoyant outflows are not directly launched by the AGN subgrid heating. }

However, following the initial heating of a central gas particle to $\deltaTagn$, it is substantially overpressurized compared to its neighbors. Consequently, it will expand rapidly and form a shock that may spread energy across a far larger region and greater mass  \citep{Springel22EntropySPH}. Such shocks will only be efficient at depositing their energy and generating entropy if they are sufficiently strong. This naturally explains the emergence of a ceiling: if gas that the AGN energy propagates through has already been sufficiently heated, it propagates as weak shocks or sound waves that dissipate minimal energy.

To show that the typical entropy predicted via this mechanism accounts for the measured ceiling value, we assume that the shock speed reaching large radii is a fraction $\zeta \lesssim 1$ of the sound speed after AGN energy deposition, 
\begin{equation}
    v_{\rm AGN} = \zeta \left(\frac{\gamma k\deltaTagn}{\mu m_p}\right)^{1/2} \simeq 1700\,\,{\rm km/s}\,\left(\frac{\deltaTagn}{1.2\times 10^8\,{\rm K}}\right)^{1/2}\zeta,\label{eq:vAGN}
\end{equation}
where the reference value of $\deltaTagn$ is adopted for the m8 Fiducial FLAMINGO volumes. 

On the upstream side of the shock one has gas with approximately the virial entropy and virial sound speed $c_s \simeq v_{\rm 200m}$, provided the shock reaches to sufficiently large radii into gas that has not yet started to cool significantly. Hence the Mach number is $\mathcal{M} \simeq v_{\rm AGN} / v_{\rm 200m}$ in the limit of a strong shock, and the Rankine-Hugoniot relations give the downstream entropy generated as
\begin{equation}
    K_{\rm shock} \simeq K_{\rm in} \mathcal{M}^2 \simeq \frac{K_{\rm 200m} }{v_{\rm 200m}^2} v_{\rm AGN}^2\,. \label{eq:Kout}
\end{equation}
Note that $K_{\rm 200m} \propto v_{\rm 200m}^2$, and therefore $K_{\rm shock}$ is constant in the strong-shock limit appropriate at small $\Mtm$. Specifically, the constant of proportionality is
\begin{equation}
\frac{K_{\rm 200m}}{v_{\rm 200m}^2} = \frac{\mu \mu_e^{2/3} m_p^{5/3}}{10(1+z)^2} \left( \frac{\pi^2 G^2} {45 H_0^4 \Omega_{\rm b,0}^2} \right)^{1/3}.\label{eq:KvirByVvir} 
\end{equation}
Combining Eq.~\eqref{eq:vAGN},~\eqref{eq:Kout} and~\eqref{eq:KvirByVvir} for the FLAMINGO cosmology gives
\begin{equation}
    K_{\rm shock} \simeq 360 \, \entropunit\,  \left(\frac{1.4}{1+z}\right)^2 \left( \frac{\deltaTagn}{1.2 \times 10^8\,{\rm K}}\right) \left(\frac{\zeta}{0.35}\right)^2\,.\label{eq:Kshock}
\end{equation}
This establishes that the entropy of $360\,\entropunit$ in Fig.~\ref{fig:entropy-inflow-outflow} can be accounted for with $\zeta \simeq 0.35$, which obeys the expected condition $\zeta\lesssim1$.  Furthermore, Eq.~\eqref{eq:Kshock} makes specific predictions for scaling with feedback strength and redshift, that we test in Sec.~\ref{sec:jets-variant}. First, however, we turn to the implications for the broader picture of how outflows escape from halos of different masses.

\subsection{Three-zone model}\label{sec:three-zone-model}
The combined shocks/buoyancy picture predicts the existence of different regimes in the stacked entropy profiles of outflowing gas:
\begin{enumerate}
\item an inner entropy-creation zone, where the mean entropy increases due to the presence of strong shocks;
\item an intermediate buoyancy zone, with constant entropy at the ceiling value, indicating that outflowing gas is being powered by buoyancy;
\item in cases where the outflow stalls, an outer heating zone representing its terminal shock. 
\end{enumerate}

\begin{figure}
    \centering
    \includegraphics[width=1.0\linewidth]{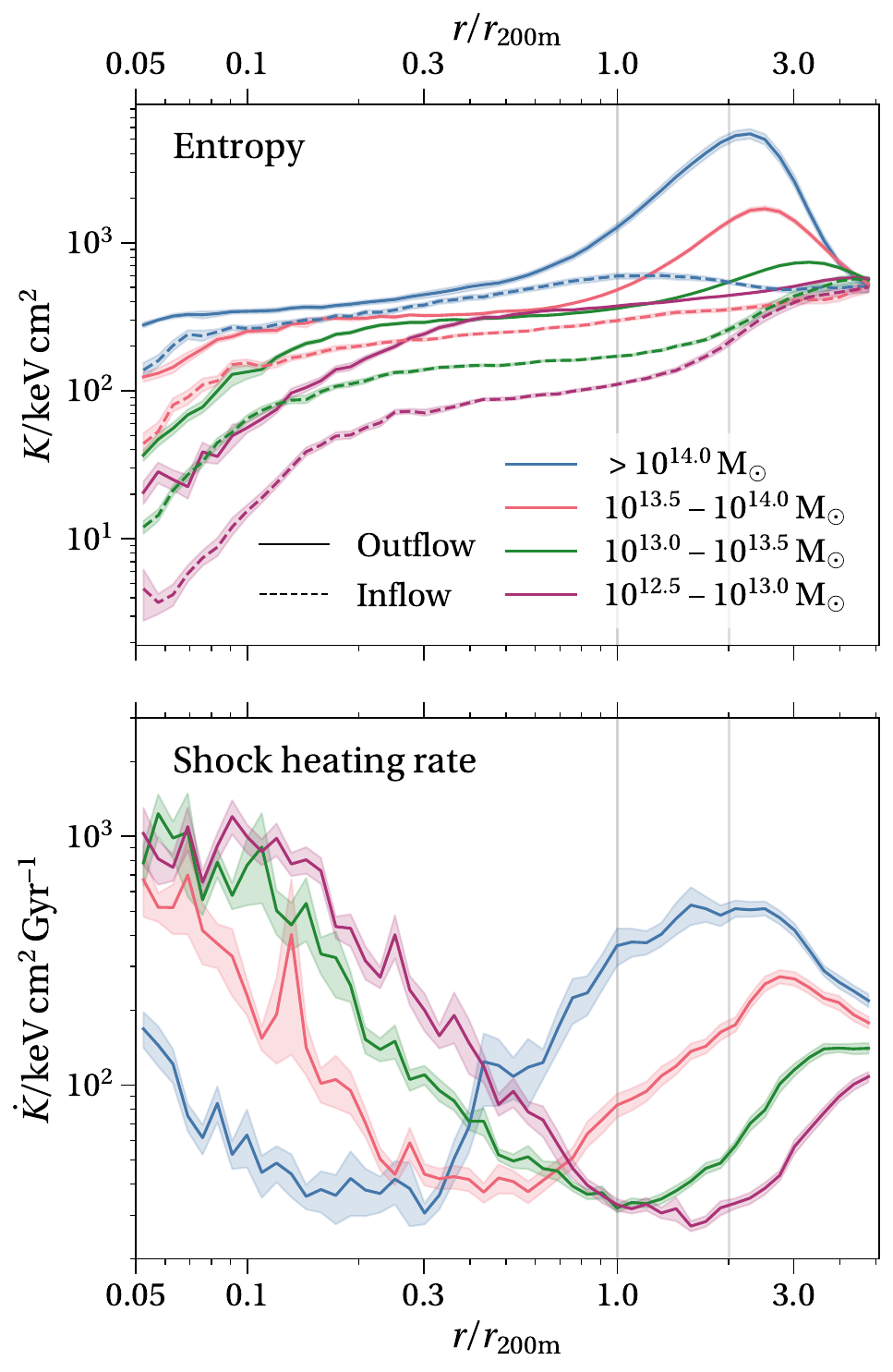}
    \caption{The entropy of inflows (dashed lines in the upper panel) and outflows (solid) as a function of normalized radius $r/\rtm$, for our chosen mass bins. The lower panel shows the mean shock-sourced entropy generation rate in each radial bin. Together these demonstrate that outflow entropy is generated by shocks in a central region ($r \lesssim 0.3 \rtm$) up to a ceiling level ($360 \,\entropunit$).  Beyond this radius, outflows maintain constant entropy until their stall radius, where a termination shock occurs. \app{Vertical lines show $\rtm$ and $2\rtm$ for convenient comparison with the splashback location (Fig.~\ref{fig:outflow-inflow-profiles}).}  } 
    \label{fig:entropy-outflow-inflow-profiles}
\end{figure}

This may be confirmed by studying the entropy as a function of radius. The upper panel of Fig.~\ref{fig:entropy-outflow-inflow-profiles} shows the radial entropy for inflows and outflows in the four mass bins previously adopted. We first briefly consider the inflows (dashed lines), which have entropies that increase with radius inside the first zone. \app{This is due to cooling in the inward-moving gas (with the effect somewhat exaggerated due to the flow-rate weighting; see Sec.~\ref{sec:post-processing}). The overall role of resolved cooling in FLAMINGO is therefore vital to supplying  gas to the central AGN, even though it has little effect on the overall energy budget (Sec.~\ref{sec:outflow-energetics}) which is dominated by the kinetic energy and by more diffuse gas.} Volume-weighted entropies (not shown here, for clarity) closely follow the outflow entropy rather than the inflow.

These outflow entropies also increase as a function of radius, as expected from the zoned model sketched above. They increase most steeply within an inner heating zone, before flattening at the entropy ceiling. The size of this zone relative to $\rtm$ shrinks as a function of mass, varying between around $0.3\rtm$ in the lowest-mass bin and $<0.05\rtm$  in the highest mass bin. This corresponds also to the heating rate $\dot{K}$, which we calculate using the simulation's artificial viscosity prescription \cite{Borrow22Sphenix}. These are presented in the lower panel of Fig.~\ref{fig:entropy-outflow-inflow-profiles}, and confirm the three-zone picture with entropy creation at small and large radii, separated by a valley that corresponds to the buoyancy zone. The variations in the size of the inner heating zone  are largely due to the variation in $\rtm$ rather than a physical difference in zone 1 between different mass halos.

The second zone, with buoyancy-driven outflows of constant entropy as a function of radius, is clearly present outside the shock zone. The extent of the zone depends on the mass of the halo. In the lowest mass bin ($10^{12.5} - 10^{13.0} \,\Msol$, purple line), it extends to several times $\rtm$, with only a small upturn in shock heating at large radii. This is because the outflow entropy far exceeds the inflow entropy, and corresponds to the behavior seen in Fig.~\ref{fig:outflow-inflow-profiles}, where gas from these halos can escape back into the Hubble flow without encountering significant resistance.  At the other extreme, in the highest mass bin ($>10^{14.0}\,\Msol$), the entropy of the outflows is never much larger than that of the inflows. Consequently, the buoyancy region terminates inside $\rtm$, and the termination-shock zone appears exactly where the outflow rates plummet (Fig.~\ref{fig:outflow-inflow-profiles}).

\begin{figure}
    \includegraphics[width=1.0\linewidth]{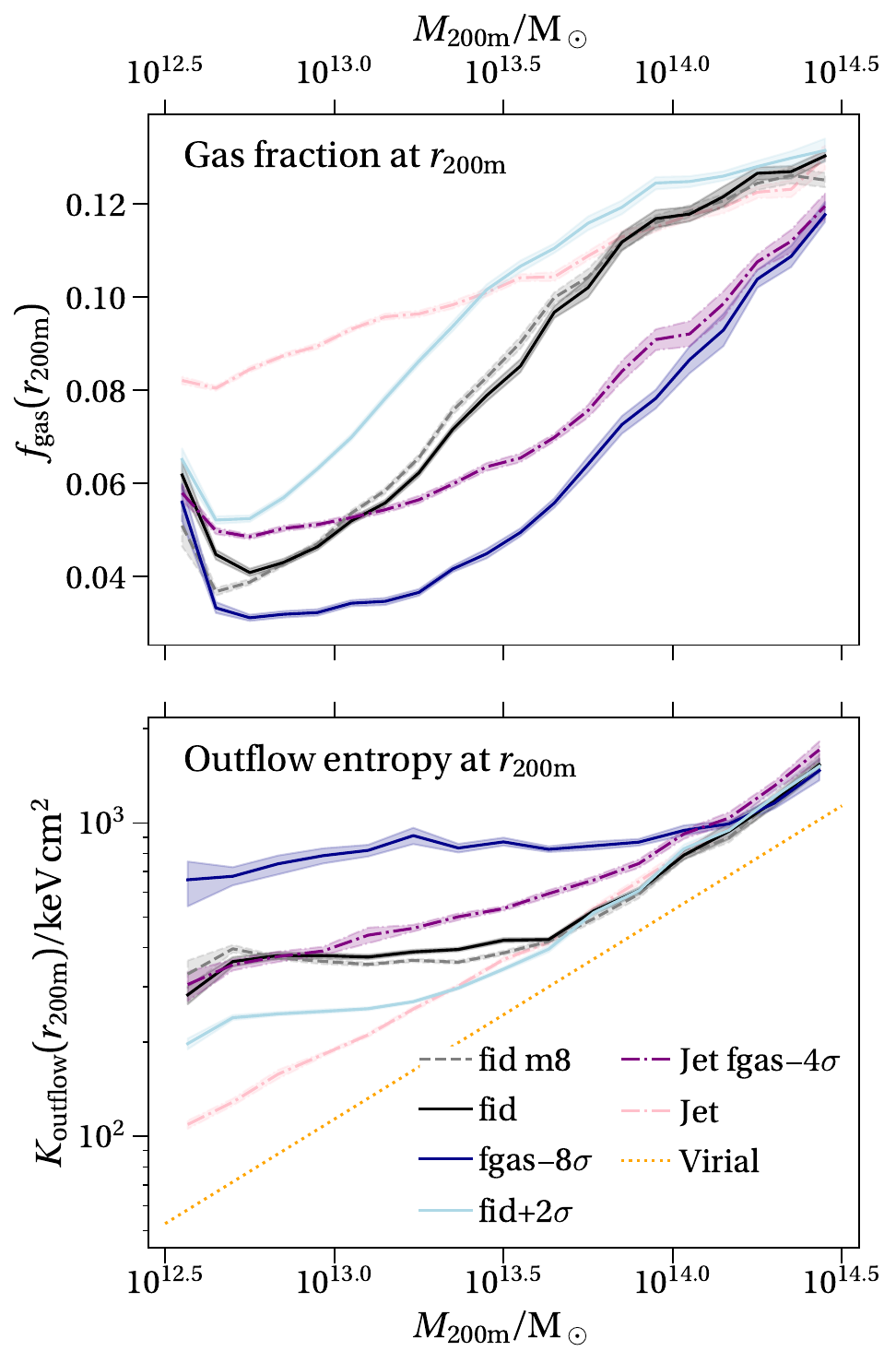}
    \caption{Summary of FLAMINGO's feedback variants in terms of enclosed gas fractions (upper panel) and outflow entropies (lower panel) at $r=\rtm$.  Dashed grey lines show the high resolution (m8) Fiducial, as used in previous figures.  Solid lines show standard resolution (m9) variants, with Fiducial, fgas$-8\sigma$ and fgas$+2\sigma$ respectively colored black, blue and light blue. Dot-dashed lines show Jet variants, with Fiducial and fgas$-4\sigma$ respectively colored light pink and purple.}\label{fig:model-comparison}
\end{figure}

\begin{figure*}
    \begin{center}
        \includegraphics[width=15cm]{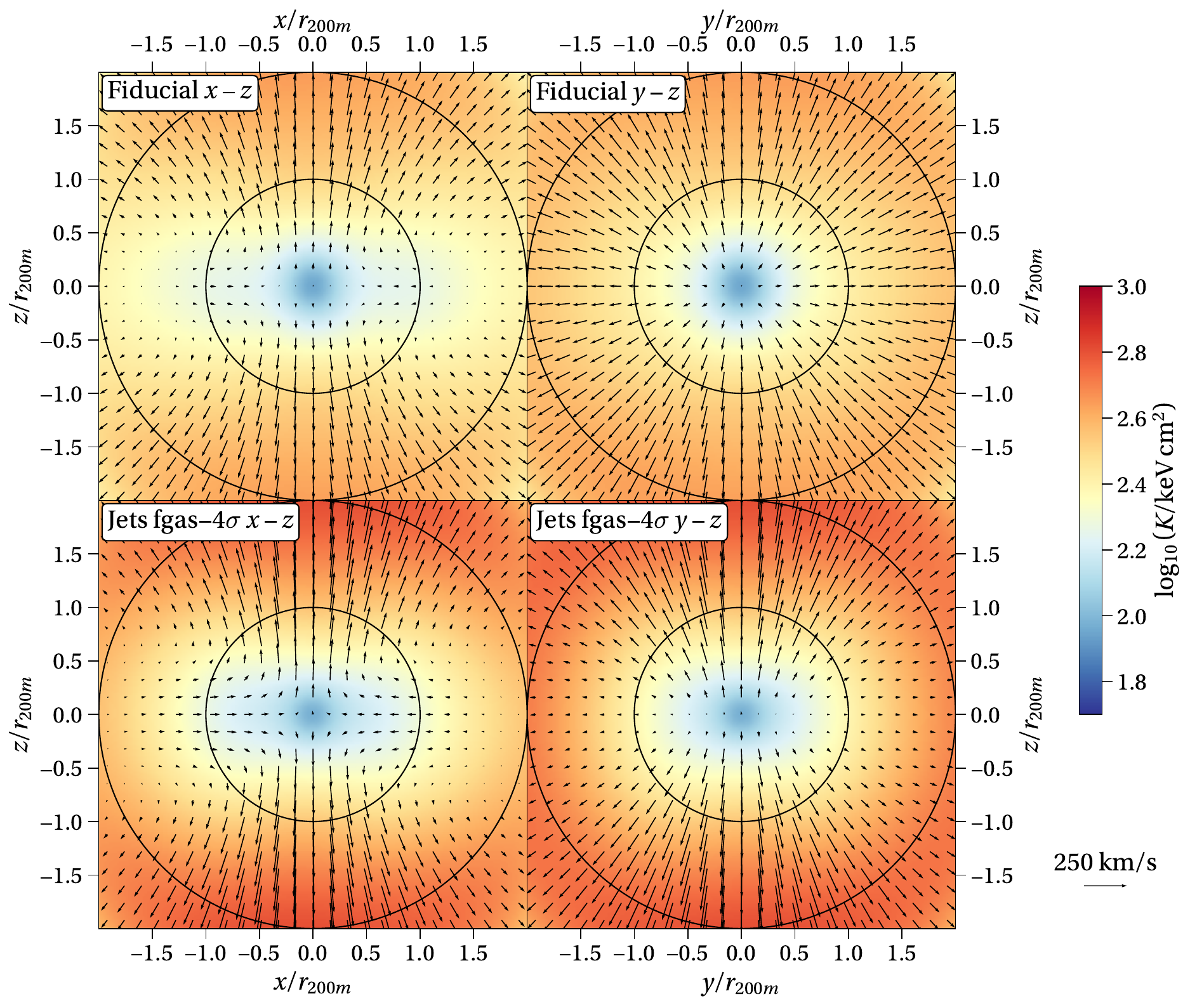}
    \end{center}
    \caption{Stacked flow and entropy plots for Fiducial (top panels) and Jet fgas$-4\sigma$ (bottom panels). Prior to stacking, halos are oriented along principal axes of the flow at the virial radius, such that the flow eigenvalues in ascending order correspond to the depicted $x$, $y$ and $z$ axes.  The mass range covered is $10^{12.5} < M/\Msol < 10^{13.0}$, a regime in which AGN-driven outflows are maximally efficient. In both feedback approaches, the low entropy of the inflow axis contrasts clearly with the high entropy of the outflow axis ($x-z$, left panels). However, in the case of the Fiducial feedback the high-entropy outflow is almost symmetric around the remaining two axes while the Jet implementation imprints a much stronger directional dependence  ($y$-$z$, right panels).  }
    \label{fig:four-panel-flow-plot}
\end{figure*}

\subsection{Feedback variants}\label{sec:jets-variant}

The shock model above predicts that the entropy ceiling should scale linearly with $\deltaTagn$ and inversely with $(1+z)^{2}$, provided that $\zeta$ is approximately constant.  This can be empirically verified using the m9 resolution feedback variants. Fig.~\ref{fig:model-comparison} compares the enclosed gas fractions $f_{\rm gas}$ (upper panel) and outflow entropies (lower panel) measured at $z=0.4$ and $\rtm$, as a function of $\Mtm$. At this radius, the m9 Fiducial model (solid black line) is well converged with the m8 Fiducial model (dashed grey line). To systematize outflow entropy measurements, we define the ceiling level as the mean outflow entropy at $\rtm$ in the mass range $10^{12.9} < \Mtm/\Msol < 10^{13.1}$. For example the m9 entropy ceiling is found at $370\,\entropunit$ compared with $360\,\entropunit$ in m8. The $f_{\rm gas}-8\sigma$ and $f_{\rm gas}+2 \sigma$ variants, which have $\deltaTagn$ respectively $2.8$ and $0.56$ times that of Fiducial, are shown by dark blue and light blue lines respectively. The scaling from Eq.~\eqref{eq:Kshock} predicts an entropy ceiling for these simulations (at $z=0.4$) of $1000\,\entropunit$ and $210\,\entropunit$ respectively, whereas empirically we find ceilings of $800\,\entropunit$ and $250\,\entropunit$, in qualitative agreement with the predicted trend.  

Similarly, the redshift scaling of Eq.~\eqref{eq:Kshock} may be tested, predicting an entropy ceiling of $180\,\entropunit$ and $730\,\entropunit$ respectively at $z=1$ and $z=0$. The measured values are $210\,\entropunit$ and $610\,\entropunit$ respectively, again in  qualitative agreement.   Given the extensive approximations being made, we take recovering these trends in a qualitative way to be a satisfactory verification of the shock-driven picture.

In addition to the variants with alternative values of $\deltaTagn$, FLAMINGO includes simulations with kinetic deposition of AGN feedback \citep{Husko22}, known as the ``Jet'' variants, which are parametrized by $v_{\rm jet}$. Results from these simulations are shown as dash-dotted lines in Fig.~\ref{fig:model-comparison}, with pink showing the Fiducial calibration and purple the fgas$-4\sigma$ variant. As measured by gas fractions, the Jet variant has far weaker outflows \app{for groups ($\Mtm \simeq 10^{13}\,\Msol$) } than the Fiducial simulation. This is because the Jet gas fractions scale more shallowly with mass, and the calibration of free parameters \citep{Kugel23} is based on X-ray data \app{with a minimum mass $M_{\rm 500c}>10^{13.5}\,\Msol$, corresponding to $\Mtm>10^{13.7}\,\Msol$}. \app{For our purpose of understanding the physics of outflows at lower masses,} the Jet fgas$-4\sigma$ simulation provides the most appropriate direct comparison to the Fiducial case.

The outflow entropy of Jet fgas-4$\sigma$, in common with the thermal cases, has a pronounced ceiling where it is elevated well above the virial entropy scaling relation (compare the purple dash-dot and orange dotted lines in Fig.~\ref{fig:model-comparison}). However, unlike the thermal cases, the ceiling is mass-dependent, albeit weakly. Fitting the mass dependence in the range $10^{12.5}\,\Msol - 10^{13.7}\,\Msol$ shows that it scales according to $K \propto \Mtm^{0.22}$. As described in our companion letter (\letter), this mass dependence arises naturally if we assume that entropy is generated solely along the jet direction, and then mixes with virial-entropy inflowing gas.

To test this picture, we produce aligned stacks of outflow speeds and entropies. For each individual halo, we first measure the flow radial velocity on a shell at $\rtm$. We  use least-squares fitting to obtain a quadratic form $\mathsf{V}$ describing that flow in terms of the unit radial vector $\hat{\vec{n}} = \vec{x}/\rtm$:
\begin{equation}
    v_r(\hat{\vec{n}}) \simeq \hat{\vec{n}}^{\top} \mathsf{V} \hat{\vec{n}}\,.\label{eq:quadratic-outflow}
\end{equation}
We then rotate the halo to diagonalize $\mathsf{V}$, such that its eigenvalues are in ascending order. We do not make further use of these eigenvalues or the approximation given by Eq.~\eqref{eq:quadratic-outflow}; its sole purpose is to select an orientation.

After rotation, the new $x$ axis corresponds to a preferred inflow direction (a negative eigenvalue); the $z$ axis corresponds to a preferred outflow direction (the largest positive eigenvalue); the $y$ axis is an intermediate axis that can exhibit either weaker outflows, weaker inflows, or approximately hydrostatic behavior. In this coordinate system, we generate individual maps of the entropy and gas velocities in a thin slice in the $x$-$z$ and $y$-$z$ planes respectively. These are stacked to produce an overall mean entropy and flow pattern. 

The results are shown in Fig.~\ref{fig:four-panel-flow-plot}. Upper panels show the m9 Fiducial simulation, while lower panels show the Jet fgas-$4\sigma$ variant. We stack in the mass range $10^{12.5} < \Mtm/\Msol < 10^{13.0}$, where \app{Fiducial and Jet fgas$-4\sigma$ have comparable gas fractions and outflow entropies (Fig.~\ref{fig:model-comparison})}. The images are scaled to have a total width of $4\rtm$, and circles indicate the radii $\rtm$ and $2\rtm$ from the halo center. Note that the stacked images must be interpreted with care, and in particular have velocity magnitudes that are much smaller than found in individual halos. Nonetheless, the stacks allow us to make a systematic comparison between the two feedback approaches.

In the $x-z$ (inflow -- outflow; left panels) projection, both variants show low entropy, gradual inflows contrasted with high entropy, rapid outflows. However, in the $y-z$ (intermediate -- outflow; right panels) projection, the differences are clear. The Fiducial model shows a near-isotropic pattern of outflows and a radial entropy gradient. The Jet variant has a much clearer distinction between a single outflow direction in which velocity is maximized and entropy increases rapidly,  while the intermediate axis is quasi-hydrostatic, with lower entropy. The stacked mean outflow velocity in the Fiducial case varies between  $70\,{\rm km\, s^{-1}}$ and $140\,{\rm km\,s^{-1}}$ (a factor $2$ variation); in the Jet case it varies between $22\,{\rm km\,s^{-1}}$ and $250\,{\rm km\,s^{-1}}$ (factor~$11$). In the case of Jet, we would therefore expect stronger spatial localisation in entropy production. Spatial variations in entropy thus generated will be diluted away as gravitational ``sloshing'' mixes material. Nonetheless, the stacked entropy varies by a factor $2$ in the Jet case, compared with $20\%$ in the Fiducial case. Overall, we take this as a satisfactory confirmation that the stronger directionality of the Jet feedback model modifies its entropy ceiling, as postulated in \letter.

\section{Discussion \& Conclusions}\label{sec:conclusions}

We have investigated why AGN feedback depletes gas in groups but is ineffective in clusters, using the FLAMINGO simulation suite. We identified buoyancy as the key physical mechanism for sustaining outflows. A ceiling to entropy creation (at $360\,\entropunit$ in the m8 Fiducial simulation) then accounts for why outflows cannot be sustained from clusters. 

Buoyant outflows are particularly energy-efficient, because gravitational dynamics transfers energy from inflowing to outflowing gas. We established the existence of this transfer through accounting of energy fluxes in Sec.~\ref{sec:outflow-energetics}. We then considered the origin of high entropy outflows, showing that these cannot be generated directly by the AGN subgrid model (which is coupled to high central densities and therefore generates only of order $40\,\entropunit$). We showed that instead the initially overpressurized gas transfers its energy outward through a shock. This gives rise to several expectations. The first is that entropy creation self-limits: once gas has been sufficiently heated, shocks will be too weak to inject further entropy, explaining the origin of the entropy ceiling. The second is a quantitative expectation for the level of entropy, based on the Rankine-Hugoniot relations (Sec.~\ref{sec:entropy-generation}; see also Ref. \cite{TangChurazov17AGNShockTheory}).  With reasonable assumptions, we showed how this accounts for the $360\,\entropunit$ simulation value, as well as its variation with feedback strength and redshift.

The final expectation is a three-zone model for gas transport in the halo, confirmed by examining the outflow entropy as a function of radius in Sec.~\ref{sec:three-zone-model}. In the innermost zone (up to $\simeq 0.3\,\rtm$), entropy is created by shocks propagating into halo gas; as radius continues to increase, this gives way to a buoyancy zone where outflow entropy plateaus. In the third zone ($r\gtrsim \rtm$), gas either rejoins the Hubble flow, stalls, or encounters a termination shock, depending on the mass of the halo. 

The transition masses between these behaviors are determined by the level of entropy created in the innermost zone.  In Sec.~\ref{sec:gas-flows}, we showed that outflows undergo a terminal shock if their entropy falls below the inflow  entropy, itself scaling with the virial value. The critical mass where net outflows therefore cease in the Fiducial FLAMINGO feedback is $\Mtm=10^{13.7}\,\Msol$.  Below this mass, outflows escape. However, where the outflow entropy is only marginally elevated, the gas lingers just beyond $\rtm$.

As the halo's virial radius grows into these stalled regions, gas fractions rise rapidly back towards the cosmic mean. This passive adjustment due to the changing virial radius causes gas fractions to increase monotonically as a function of mass for $\Mtm>10^{13.0}\,\Msol$. We demonstrated the connection between lingering gas and rising gas fractions by deriving and evaluating a three-term expression for $\dot{f}_{\rm gas}$ (Sec.~\ref{sec:fgas-non-gas}). Of the three terms, only one is directly related to net gas flow across the virial radius; the other two correspond to total mass flow (dominated by dark matter), and the expansion of the virial radius including its pseudoevolution. 

\app{In FLAMINGO, strong shocks are the mechanism for generating outflow entropy, by coupling AGN energy to a large volume of low-density gas. Sec.~\ref{sec:entropy-generation} connected the subgrid parameter $\Delta T_{\rm AGN}$ with the ceiling entropy, with Sec.~\ref{sec:jets-variant} showing the characterization correctly captures variation with feedback mechanism, strength and cosmic time. In turn, $\Delta T_{\rm AGN}$ is set by calibration to X-ray observations \cite{Kugel23};  in a loose sense, therefore, the entropy ceiling is connected to an observational calibration rather than a subgrid parameter. }

We showed that the overall picture continues to apply in the alternative subgrid description offered by the Jet simulations (Sec.~\ref{sec:jets-variant}). Moreover, where the virial $\fgas$ values agree, so too do the outflow entropy values (Fig.~\ref{fig:model-comparison}). However the quantitative scaling of outflow entropy with mass does depend on feedback geometry. To show this, we introduced a simple method for generating oriented stacks of anisotropic flows (Sec.~\ref{sec:jets-variant}). We demonstrated that in Fiducial (thermal) implementations, AGN-driven outflows propagate along two axes, with the third being dominated by inflows. Conversely, in the Jet (kinetic) AGN variant, outflows are oriented along a single axis, with near-hydrostatic behavior on the second axis.  The more tightly collimated outflows give rise to an entropy ceiling that increases modestly with mass ($K\propto \Mtm^{0.22}$), a phenomenon that is explored in \letter. However, the virial entropy scales more steeply with mass ($\propto \Mtm^{2/3}$), and therefore buoyancy still fails at a critical mass determined by the strength of the kinetic outflows. 

\app{There is no suggestion that either of the subgrid descriptions captures the microphysics of AGN. Boiled down to its minimum, the proposal here is quite general: AGN feedback can initially be deposited in any way, provided it later thermalizes in low-density gas through a self-limiting process}. The importance of reaching low densities is made clear by considering the scaling of entropy $K$ in terms of internal specific energy $u$ and density $\rho$ (i.e. $K \propto u/\rho^{2/3}$). To achieve outflows that are buoyant all the way to the virial radius and beyond, it is not sufficient to elevate $u$ to some multiple of the virial temperature. Rather, this heating must occur at densities $\rho$ that are comparable with the virial density, given by
\begin{equation}
    \rho_{\rm b, 200m} = 200\,\Omega_{b,0} (1+z)^3 \rho_{\rm crit,0} \simeq 5 \times 10^{-5}(1+z)^3\,m_p\,{\rm cm}^{-3}\,.
\end{equation}
Observed densities decline steeply to this value between around 20\% and 50\% of $r_{\rm 500c}$, equivalent to between 10\% and 50\% of $\rtm$ \cite{Sun12GasDensity,Eckert12GasDistribution,GhirardiniICMprofiles19,Pratt22DensityEstimates}. If the energy from the AGN successfully thermalizes in this far-out, tenuous gas, it will enable a buoyant outflow. \app{On the other hand, if the tenuous gas is already hot, it becomes intrinsically hard to pump energy into it non-adiabatically due to the high sound speed, generating the self-limiting `thermostat' effect. }

\app{Given this expectation, it is of particular interest to compare our results with those from} Illustris, SIMBA and XFABLE, which all decouple feedback from the hydrodynamics until it reaches large halo-centric radii. \app{Their approach} has been found to be particularly efficient at depleting baryons in groups \cite{Lim2021IllustrisGasFrac,Sorini22SIMBA,Ayromlou23ClosureRadius,Bigwood25XFABLE}. The buoyancy-driven picture explains this result: these schemes naturally distribute energy into tenuous gas, achieving a high entropy ceiling. The same effect is achieved by FLAMINGO's fgas$-8\sigma$ variant which increases $\Delta T_{\mathrm{AGN}}$ and hence enables shocks to propagate to larger radius, where the higher entropy is naturally produced.  In both cases one expects a higher entropy ceiling, and therefore outflows from halos of higher masses.

Most contemporary cosmological simulations, however, share a common limitation in omitting sources of non-thermal pressure such as magnetism or cosmic rays   \cite{Sijacki08AGNCR,Ehlert18MagneticJetsCRsClusters,Ruszkowski23CRreview,Grete25IdealisedMHDGroupsClusters,Meenakshi26ClusterHeating}. Recent observational evidence  in the central $\sim 100\,\kpc$ of clusters suggests subgrid models produce velocity dispersions that are $50$--$70\%$ too large  \cite{XIRSM25VelDispersions}, underscoring the need to examine alternative energy transport on small scales. Cosmic rays can directly add buoyancy to bubbles, confined by magnetic fields, while plasma transport processes (anisotropic viscosity, conduction and streaming) alongside turbulent dissipation can compete with shocks as entropy sources. Studies on such microphysical mechanisms have largely been directed at the problem of regulating cooling flows in massive clusters, and it would be of significant interest to probe the interaction with the outer halo in the group-scale regime where the FLAMINGO entropy ceiling is found. The essential ingredients of buoyancy breakdown at a critical mass, of a mediation zone between microphysics and buoyant transport, and of gas lingering and reincorporation around the splashback radius, will plausibly survive even with these complications. 

Turning to observational constraints on halo scales, joint weak-lensing, kSZ, and X-ray analyses of DESI/ACT and eROSITA data \cite{Bigwood24KSZ, McCarthy25kSZSims, Hadzhiyska25kSZDESI, Siegel25JointConstraints} are compatible with the picture presented here. In particular, they are consistent with a strong mass dependence of the gas fraction \cite{LLS25} and with recovering the cosmic baryon budget within $2-3 \rtm$ of group-scale halos in kSZ+CMB-lensing measurements \cite{Hadzhiyska25MissingBaryonsKsz}. This matches the lingering-and-reincorporation phenomenology that we have shown results naturally from buoyantly-driven outflows. The increasing precision of tSZ and kSZ constraints from Simons Observatory \cite{Ade2019SO}, and dispersion measures from FRBs \cite{Wayland26FRB}, will tighten these constraints in the imminent future.  

A complementary, galaxy-scale view is examined in \letter. There, we confront the model with new evidence from population-level stellar mass and star formation rate inference using \textsc{pop-cosmos} \cite{alsing2024,thorp2025}, combined with JWST NIRCam morphologies from COSMOS-Web \cite{shuntov25cosmosweb}. We discuss how the most massive low-redshift galaxies exhibit declining quenched fractions as a function of time \cite{deger2025} and a broad range of structural types. This hints at the onset of rejuvenation in massive central galaxies. 

\app{Altogether, there is therefore growing observational evidence for the existence of a limit to AGN effectiveness, for which buoyancy breakdown is a plausible explanation.}    Ideally, one would like to find observational evidence for shock heating as a mediator. The majority of observationally-detected AGN-driven shocks are weak and therefore carry energy losslessly, generating little entropy \cite{McNamara05ClusterShockHeating,Fabian06Perseus,Forman07M87Shocks,Randall15NGC5813Shocks,Liu19AGNShock,Ubertosi23AGNShock,Omoruyi26AGNShock}. This is consistent \app{with the hypothesized existence of strong shocks in groups and sub-groups} because the observational sample is strongly biased towards high-mass clusters. A rare example of an unambiguous strong AGN-driven shock has been detected in Centaurus A \citep{Croston09CenAShock}, a sub-group with $\Mtm \simeq 10^{12.5} \,\Msol$. However, this is found in bright, high-density gas a few kpc from the center, and therefore unlikely to be a source of a high entropy halo-scale outflow. The deepest X-ray observations such as those in Ref. \cite{Sun09XrayChandraGroupsClusters} reach ${\sim}r_{500\mathrm{c}}\simeq 0.6\,r_{200\mathrm{m}}$, but only when averaged in circular bins, and with $\sim 100\,\mathrm{eV}$ spectral resolution. Consequently it is very hard to resolve shocks or separate the
diffuse outflowing phase from denser inflowing gas.

To reach larger radii and the low-mass regime, future X-ray missions will need a wide field of view, sensitivity to soft, low surface brightness emission, and high spectral resolution \cite{Zhang24HighResXraySpec}. A group at $M_{200\mathrm{m}} \simeq
10^{13}\,M_\odot$ and $z \simeq 0.2$ has a radial extent of 
$\simeq 3\,\mathrm{arcmin}$, containing gas at $kT \sim
0.5\mathrm{keV}$ and $n_e \simeq 5 \times
10^{-5}\,\mathrm{cm}^{-3}$. 
The forthcoming NewAthena \cite{Cruise25NewAthena} and, especially, the proposed LEM (Line Emission Mapper; \cite{Kraft22LEM}) push towards the required regime to map shocks and outflows.  In particular LEM, with $15\, \mathrm{arcsec}$ angular resolution, $1$--$2\,\mathrm{eV}$ spectral resolution, $>1500\,\mathrm{cm}^2$ effective area, and an ample $30\,\mathrm{arcmin}$ field of view, may be able to map and separate inflow and outflow phases. This would allow for testing of our key broader prediction: that mean entropy of outflows at $\rtm$ will be shallow or flat as a function of mass between sub-group and sub-cluster scales.

\section*{Acknowledgements}

We thank Alex Amon, Leah Bigwood, Andy Fabian, George Efstathiou, Simone Ferraro, Boryana Hadzhiyska, Filip Huško, Cedric Lacey, Luisa Lucie-Smith, Ian McCarthy, Daisuke Nagai, Emmanuel Schaan and Tom Theuns for helpful conversations.  This work has been supported by funding from the European Research Council (ERC) under the European Union's Horizon 2020 research and innovation programmes (grant agreement no.\ 818085 GMGalaxies and no.\ 101018897 CosmicExplorer), and the research project grant ``Understanding the Dynamic Universe'' funded by the Knut and Alice Wallenberg Foundation under Dnr KAW 2018.0067. HVP was additionally supported by the G\"{o}ran Gustafsson Foundation for Research in Natural Sciences and Medicine. AP and HVP thank the organizers of the inspiring {\it Cosmic Ecosystems} workshop held in July 2025 at the Perimeter Institute, where this work was initiated, for their kind hospitality.  This research was supported in part by Perimeter Institute for Theoretical Physics. Research at Perimeter Institute is supported by the Government of Canada through the Department of Innovation, Science, and Economic Development, and by the Province of Ontario through the Ministry of Colleges and Universities.

\section*{Author Contributions}
{\bf AP}: conceptualization; formal analysis; investigation; methodology; software; validation; visualization; writing -- original draft, review \& editing.
{\bf HVP}: conceptualization; formal analysis; investigation; methodology; validation; visualization; writing -- review \& editing. 
{\bf JS}: Project administration (FLAMINGO); data curation; resources; writing -- Review \& Editing.
{\bf MS}: Data curation; Resources.

\section*{Declaration of LLM use}

We used Anthropic's Claude Opus 4.6 to aid in code development, to proof-read text and formulae, and to search existing literature.

\bibliography{example}

\label{lastpage}
\end{document}